\documentclass[10pt,letterpaper,compsoc,conference]{iiswc23}

\usepackage{url}
\usepackage{graphicx}
\usepackage{amsmath,amssymb}
\usepackage[colorlinks = true,
            linkcolor = blue,
            urlcolor  = blue,
            citecolor = red,
            anchorcolor = blue]{hyperref}
\usepackage{listings}
\usepackage{tabularx}
\usepackage{authblk}
\usepackage{listings}
\usepackage{mathtools}
\usepackage{graphicx}
\usepackage{framed}
\usepackage{comment}
\usepackage{listings,newtxtt}
\usepackage{multirow}
\usepackage{dblfloatfix}    
\usepackage[table]{xcolor}

\usepackage[dvipsnames]{xcolor}
\usepackage{enumitem}
\usepackage{tikz}
\usepackage{xcolor}

\usepackage{circledsteps}
\usepackage{pbox}
\usepackage{comment}
\usepackage{csvsimple}
\usepackage{fancyhdr}
\usepackage{mathtools}
\usepackage{caption}
\usepackage{amsmath}
\usepackage{makecell}
\usepackage{tikz}
\usepackage{subfigure}
\usepackage{pifont}
\usepackage{color}
\usepackage{xspace}

\usepackage{siunitx}
\pgfkeys{/csteps/inner color=white}
\pgfkeys{/csteps/fill color=black}

\definecolor{codegreen}{rgb}{0,0.6,0}
\definecolor{codegray}{rgb}{0.5,0.5,0.5}
\definecolor{codepurple}{rgb}{0.58,0,0.82}
\definecolor{backcolour}{rgb}{0.95,0.95,0.92}
\lstdefinestyle{mystyle}{
    backgroundcolor=\color{backcolour},
    commentstyle=\color{codegreen},
    keywordstyle=\color{magenta},
    numberstyle=\tiny\color{codegray},
    stringstyle=\color{codepurple},
    basicstyle=\ttfamily\footnotesize,
    breakatwhitespace=false,
    breaklines=true,
    captionpos=b,
    keepspaces=true,
    numbers=left,
    numbersep=5pt,
    showspaces=false,
    showstringspaces=false,
    showtabs=false,
    tabsize=2
}
\newcommand{\cmark}{\textcolor{ForestGreen}{\ding{51}}}%
\newcommand{\xmark}{\textcolor{red}{\ding{55}}}%

\begin{document}

\newcommand{\gem}{gem{5}}
\newcommand{\papertitle}{HammerSim\xspace}

\title{\papertitle: A System-Level Tool to Model RowHammer} 


\author{Kaustav Goswami}
\author{Ayaz Akram}
\author{Hari Venugopalan}
\author{Jason Lowe-Power}
\affil{University of California, Davis}

\maketitle
\pagestyle{plain}

\begin{abstract}

  Modern architecture research relies on simulators to evaluate system security, yet analyzing emerging hardware vulnerabilities like RowHammer requires full-system visibility.
  As RowHammer vulnerabilities worsen with continuous technology scaling, existing simulators lack the system-level models needed to study complex OS effects and cross-layer mitigations.
  This tool deficiency leaves modern computing platforms exposed to severe reliability and security risks.
  In this work, we present \papertitle, a \gem-based framework for modeling RowHammer at the full-system level.
  \papertitle integrates probability-driven bitflip modeling to realistically capture the behavior of RowHammer.
  It further enables evaluation of hardware and software mitigations such as TRR and selective ECC.
  We validate \papertitle's bitflip modeling against real DDR4 DIMMs using JS divergence, demonstrating its utility in studying attacks, defenses, and benign workload susceptibility.
  Our framework provides an extensible platform to bridge the gap between hardware experiments and architectural simulation.

\end{abstract}


\section{Introduction}
    \label{sec:label}
    Dynamic random access memory, or DRAM, is the \textit{de-facto} choice for main memory design due to its cost-effectiveness.
    However, due to scaling, researchers have noticed variations in the nominal parameters of these devices~\cite{variation_paper_1, process, moesi_rh}.
    Recently, it has been found that accessing (or \texttt{ACT}ivating) the same cell or the same set of cells on a DRAM module repeatedly causes data corruption in the neighboring cells.
    This is called RowHammer~\cite{intel-patent, para}.
    Researchers have shown several system-level attacks using RowHammer to escalate privileges, fingerprint or leak private data~\cite{fphammer,centauri,rowhammer-puf,rowhammer-leak,seaborn2015exploiting,orosa2022spyhammer,rowhammerjs,spechammer,anvil,windows_rowhammer,ecc_rowhammer,cloud-attack,gpu-herbert-bos,trrespass,yet-another-flip,sgx-bomb}.

    Currently, researchers rely on two primary hardware-based methods to study RowHammer: \textit{(a)} FPGAs, and \textit{(b)} software-based exploits~\cite{characterization_paper, rh_micro_paper, hammertime, ddr_time}.
    FPGA-based memory controllers~\cite{hassan2017softmc, retro_rh, drambender} (SoftMC~\cite{hassan2017softmc}, DRAM-Bender~\cite{drambender}, \textit{etc.}) provide highly accurate timing and have been instrumental in uncovering proprietary in-DRAM mitigations like Target Row Refresh (TRR)~\cite{utrr}.
    However, FPGAs lack the capability to model OS-level abstractions or full server-class system dynamics. 
    Conversely, software-based RowHammer exploits~\cite{trrespass,blacksmith,smash,spechammer,sledgehammer,rowhammerjs,drammer} successfully bypass OS-level protections and allow researchers to explore kernel-level vulnerabilities~\cite{anvil,fredchongkernelpaper}.
    Yet, they do not provide the hardware visibility necessary to prototype or evaluate new hardware-software co-design mitigations.



    Researchers today use cycle-level simulations~\cite{para,blockhammer,prohit,graphene,twice,cratar,cbt} to bridge this gap between the hardware-level and software-level view of RowHammer.
    However, popular cycle-accurate and cycle-level simulators~\cite{dramsim2,dramsim3,ramulator,lowe2020gem5,nvmain} used for memory system research today assume an idealistic setting, often using timing, current, and voltage values from data sheets for simulating hardware components~\cite{wang2005dramsim, dramsim2, dramsim3, ramulator, nvmain, lowe2020gem5}.
    To model variations, researchers have to explicitly add variations~\cite{vampire, process, memsys-crosstalk}, which generally leads to over/under-estimation of the simulation model.
    While some models of process variation exist purely as a statistical models~\cite{varius,delft,memsys-crosstalk}, we do not have models for RowHammer.
    Further, the RowHammer threshold is decreasing as DRAMs are becoming denser~\cite{kim2020revisiting}, which motivates the need for such a model.

\begin{figure}[t]
        \centering
        \includegraphics[width=0.45\textwidth]{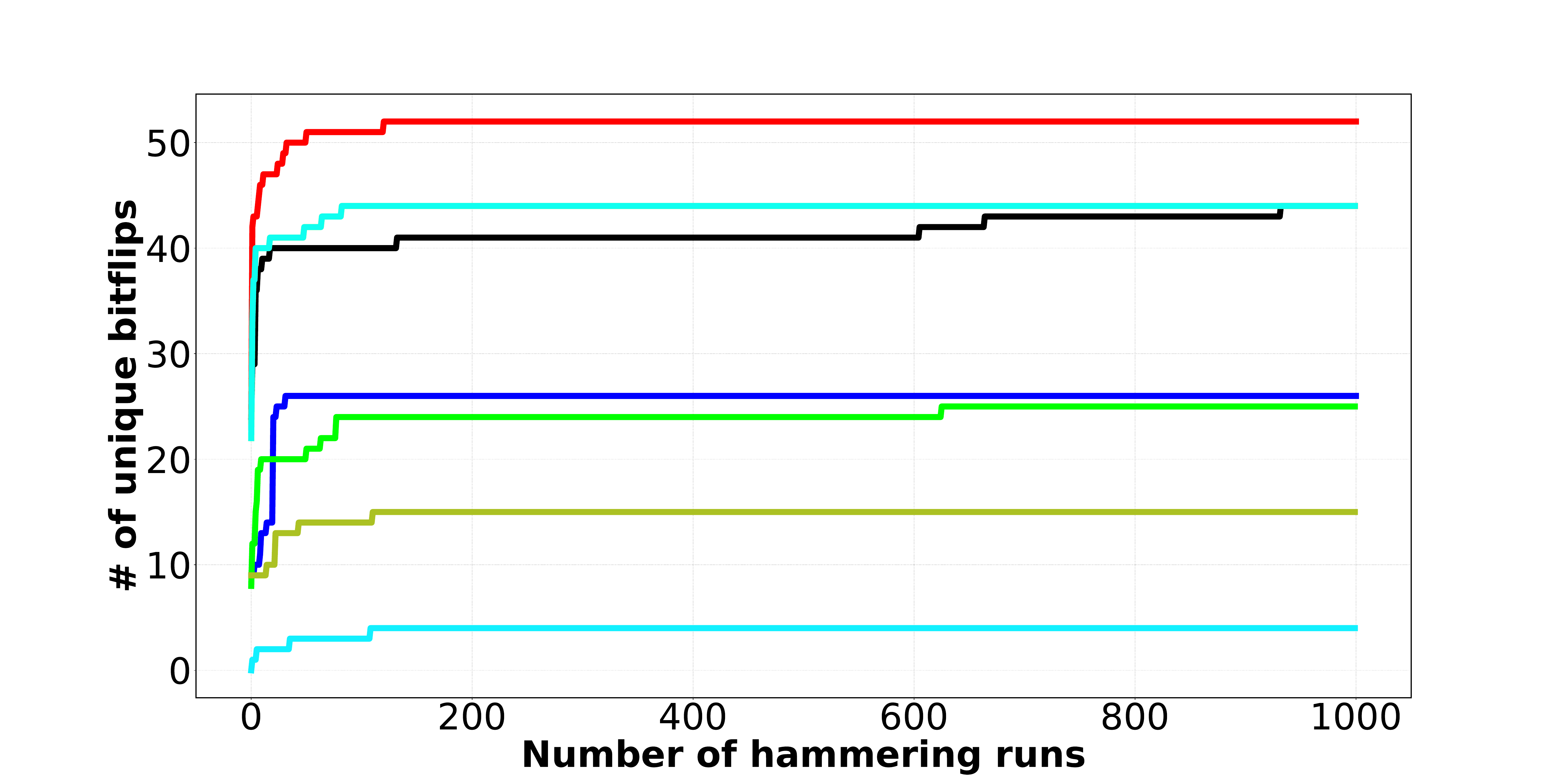}
        \caption{Count of unique bitflips on different DRAM DIMMs' row. Each
        line plot corresponds to a victim row. In total, we collected victim
        row data from 9 different DIMMs from two different manufacturers.}
        \label{fig:rh-count}
    \end{figure}

    To understand the fidelity gap of such uniform models, we performed extensive RowHammer characterization experiments across multiple real-world DRAM Dual Inline Memory Modules (DIMMs).
    As shown in Figure~\ref{fig:rh-count}, our hardware evaluations reveal that RowHammer-induced bitflips are highly non-uniform; unique bitflips saturate over time, and certain memory regions exhibit significantly higher vulnerability than others.

    Crucially, prior literature demonstrates that real-world RowHammer behavior is deeply intertwined with full-system parameters, including the operating system page allocation~\cite{windows_rowhammer,seaborn2015exploiting,blacksmith}, CPU frequency~\cite{centauri}, cache coherency protocols~\cite{moesi_rh}, instruction set architectures (ISAs)~\cite{para,seaborn2015exploiting,trrespass,blacksmith,anvil,drammer,riscv-rowhammer}, and underlying memory technology~\cite{para,twice,utrr}. 
    Even benign applications running under normal system workloads have been shown to inadvertently trigger RowHammer bitflips~\cite{abacus}.
    Without the ability to simulate how hardware faults dynamically interact with OS scheduling and cache hierarchies, researchers cannot reliably estimate security boundaries or system reliability.

    To solve this, we propose \papertitle, a full-system RowHammer modeling framework built upon the gem5 architectural simulator~\cite{lowe2020gem5}.
    \papertitle directly addresses the shortcomings of prior tools by combining the physical accuracy of hardware-characterization maps with the system-level visibility of an architectural simulator.
    While circuit-level models~\cite{circuit_paper} or trace-driven simulators cannot natively capture live OS reactions, \papertitle runs an unmodified OS.
    It enables researchers to analyze realistic RowHammer data corruption under complex workloads, cache hierarchies, and OS scheduling, while actively supporting the evaluation of sophisticated hardware-software mitigation co-designs


    

    Overall, the contributions of this paper are as follows:
    \begin{itemize}
        \item System-level RowHammer model in gem5 that incorporates multiple probability distributions for realistic bitflip estimation, enabling studies on real workloads and future DRAM technologies with or without mitigations.
        \item Dual analysis modes with data corruption support where an \textit{online mode} enables in-situ data corruption during simulation to study system-level effects, and an \textit{offline mode} provides post-processing for accurate probability estimation and correction of over/under-estimations.
        \item Hardware-informed modeling and validation where extensive RowHammer experiments on DDR4 DIMMs to capture real-world bitflip patterns and validate simulation fidelity using JS divergence.
    \end{itemize}

    The rest of the paper is organized as follows. Section~\ref{sec:background} discusses the basics of a DRAM device, the RowHammer vulnerability and 
    Section~\ref{sec:method} describes our RowHammer model in a hollistic manner.
    Section~\ref{sec:results} analyzes the simulations.
    The work in concluded in Section~\ref{sec:conclusion} with further scope of improvement.
\section{Background}
    \label{sec:background}

    In this section, we introduce the basics of a DRAM device, how it is
    implemented as an interface in \gem, and the RowHammer vulnerability.

    \begin{figure}[h]
    \centering
      \includegraphics[width=\columnwidth]{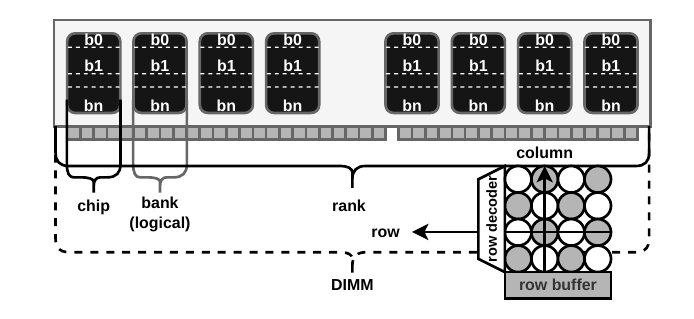}
      \caption{This figure shows a single rank of a DRAM DIMM. Each rank contains 
                multiple logical structures called banks that are interspersed 
                across multiple physical structures called chips.}
      \label{fig:dram_arch}
    \end{figure}

    \subsection{DRAM Organization} 
        \label{subsec:organization}
        DRAM has spawned several generations and technologies.
        This includes DDR\texttt{x}~\cite{ddr2,ddr3,ddr4,ddr5}, GDDR\texttt{x}~\cite{gddr5,gddr6}, LPDDR\texttt{x}~\cite{microLPDDRx}, DDR\texttt{x}L~\cite{ddr3l}, \textit{etc.} 
        Each physical DRAM Dual Inline Memory Module or DRAM DIMM is installed on a DRAM \textit{channel} on the motherboard. 
        Channels permit issuing concurrent memory requests to multiple DIMMs. 
        A DIMM contains multiple logical structures called \textit{banks} on one or both of the \textit{ranks}. A bank is a two-dimensional array of cells organized into \textit{rows} and \textit{columns}.
        Each cell contains a capacitor and an access transistor, with the capacitor's charged state representing a single bit. Organizing cells into banks enables faster memory access since memory addresses that map to different banks can be accessed in parallel.
        Independent dies are packaged into multiple physical entities called \textit{chips}.
        The banks on each rank are uniformly distributed across these chips (Figure~\ref{fig:dram_arch}).
        



    \subsection{Working of a DRAM Device}
        \label{subsec:working-dram}
        
        \label{subsubsec:mem_req}
        The memory controller receives memory requests with an instruction and a corresponding address to the DRAM.
        These addresses encode information about the DRAM structure (channel, rank, bank, etc). A mapping function uniquely maps addresses to a set of capacitors within the DRAM~\cite{drama}\footnote{Note that this is not a virtual-to-physical address mapping. The virtual address is translated to its corresponding physical address in the Translation Lookaside Buffer (TLB)~\cite{hennesyandpatterson}. This mapping is physical to device mapping.}.
        On receiving a memory request, the memory controller uses this function to identify the corresponding bank within the corresponding DIMM. The address is then pushed into the bank's \textit{row-decoder} logic.
        Then the row-decoder selects the appropriate row and transfers it to a \textit{row-buffer}, which is known as an \texttt{ACTIVATE}, or \texttt{ACT}.
        The row-buffer is an array of sense amplifiers that sense the charge of the cells in that row to determine their bit values.
        Finally, the bank's \textit{column-decoder} reads the values of the appropriate cells within the row-buffer and returns the corresponding data.

        \label{subsubsec:ref}
        DRAM technology is based on capacitors, which inherently lose its charge over time~\cite{memorybook}. 
        Their contents have to be recharged periodically to preserve the integrity of the data stored. The memory controller issues a \texttt{REFRESH (REFI)} instruction to a subset of rows every 7.8 $\mu$s (\texttt{tREFI}) to preserve the values of their capacitors. 
        To refresh a particular row, it is first activated to sense the bit values.
        Then, the bits are subsequently restored to the DRAM array.
        Every DRAM capacitor gets refreshed at least once every 32 ms or 64 ms (\texttt{tREFW}) based on the operating
        temperature.

    \subsection{\gem's Memory Controller and Memory Interface}
        \label{subsec:gem5_memctrl}
        The memory controller in \gem~is simulated as an event-driven component~\cite{gem5_memctrl_paper}.
        It models a generic memory controller with split queues for reads and writes, and a shared queue for responses.
        It is responsible for managing the communication between the processor and the memory.
        gem5 receives memory requests from the processor and schedules these requests based on a given scheduling policy.
        The memory organization it models is similar to its regular counterpart.
        Addresses decode into rank, bank, row, and column at each memory controller.
        The channel interleaving is decoded outside at a separate crossbar.
        A Packet is used to encapsulate a transfer between two objects in the memory system~\cite{lowe2020gem5}.
        An LLC miss in \gem~is seen as a packet coming into the memory controller.
        Once the packet is received at the memory controller, it decodes that packet, schedules it, and sends it to the memory interface.


        Memory devices are modeled as classes in \gem.
        Currently in \gem, there are two memory interfaces: (a) DRAM and (b) NVM.
        The timing parameters for these devices are based on the technical sheet of these devices.
        Hansson \textit{et al.}~\cite{gem5_memctrl_paper} mentions that \gem~does not model all the timing parameters from a contemporary memory device.
        However, gem5 models all the important timing parameters, which were validated against a cycle-accurate DRAM simulator.            

    \subsection{The RowHammer Vulnerability}
        \label{subsec:rh}
        Scaling of DRAM devices over the generations has resulted in the deviation of their nominal parameters from their theoretical counterpart due to variations~\cite{process,variation_paper_1,rowhammer-puf}. 
        Modern DIMMs are susceptible to memory corruption as a result of electrical interference among the cells~\cite{rowpress}.
        One such technique to leverage bit flips without accessing the row is the rowhammer (RH) exploit~\cite{intel-patent, para}.
        It causes memory corruption by repeatedly accessing two addresses that map to two different rows in a single bank.
        This leads to the repeated activation and deactivation of the two rows back-and-forth between the row-buffer and the 
        DRAM bank array.
        The resulting electro-magnetic interference between the accessed rows (referred to as \textit{aggressor rows} henceforth) and their neighboring rows (referred to as \textit{victim rows} henceforth) accelerates the rate at which the capacitors in the victim rows lose their charge.
        Once capacitors have lost a sufficient amount of charge, refreshing the DRAM does not restore their value, resulting in memory corruption in the form of bit flips.


        There have been several works done in the past to detect and mitigate RH attacks~\cite{para, prohit, cbt, twice, graphene, blockhammer}. 
        DRAM manufacturers implement TRR added the standard mitigation mechanism in DDR4 devices as a practical solution that does not violate DRAM timings~\cite{ddr4-samsung,ddr4,trrespass}.
\section{\papertitle}
    \label{sec:method}

    Our objective is to develop a comprehensive, full-system simulation framework that accurately mirrors the hardware-level probabilistic and spatial characteristics of RowHammer.
    In this section, we list the challenges, perform extensive hardware exploration, incorporate the findings and discuss the implementation details.

    \begin{figure}[t]
        \centering
        \includegraphics[width=0.4\textwidth]{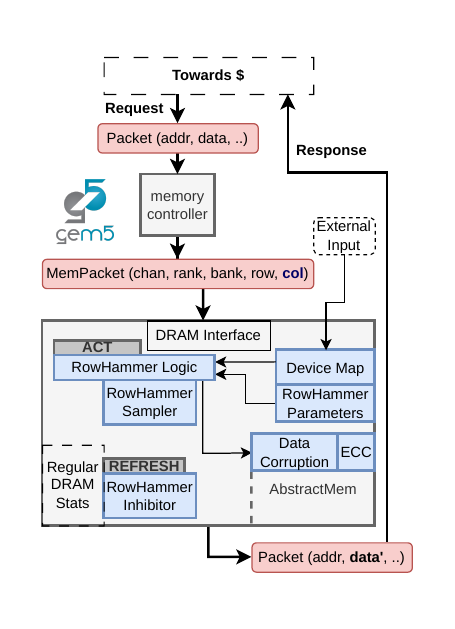}
        \caption{High-level overview of RowHammer Modeling in \gem.
            \papertitle architectural components are highlighted in blue.
            The core modeling logic is integrated within the \gem~DRAM interface, while the execution path is configured via flexible user parameters.}
        \label{fig:rh}
    \end{figure} 
    \subsection{Architectural and Simulation Challenges}
        Standard circuit-level or trace-driven memory simulators cannot resolve the three fundamental design challenges introduced when designing a scalable, full-system simulation infrastructure for RowHammer:

        \begin{enumerate}[label=C\arabic*]
            \item \textbf{Representing Non-Uniform Hardware Non-Idealities.} Traditional architectural simulators treat faults as uniform or binary events.
            However, real-world hardware data reveals that RowHammer bitflips are heavily non-uniform and highly localized due to manufacturing process variations.
            Translating these irregular physical vulnerabilities into deterministic, flexible, and mathematically sound probability models within an architectural simulator is highly non-trivial.
            \item \textbf{Spatial Tracking Overhead at Scale.} To model multi-row vulnerabilities (e.g., Half-Double~\cite{halfdouble,half-double-paper} or multi-sided attacks~\cite{para,trrespass}), the memory controller must continuously maintain and update activation tracking structures across spatial neighborhoods.
            Performing these multi-row spatial lookups at runtime for every memory access introduces severe simulation timing and memory overheads, requiring highly optimized metadata tracking.
            \item \textbf{Balancing Execution Fidelity with Multi-Parameter Exploration.} To evaluate hardware-software defenses, researchers require an \textit{online mode} capable of dynamically corrupting live memory data to capture operating system reactions and application-level crashes.
            Conversely, security researchers exploring different hardware topologies require rapid, trace-driven sensitivity sweeps.
            The framework must seamlessly decouple full-system execution from trace-driven analytical processing without duplicating simulator infrastructure.
        \end{enumerate}

    \begin{table}[t]
        \caption{Experimental System Parameters}
        \label{tab:kaby-lake}
        \centering
        \begin{tabular}{cc}
            \hline
            \textbf{Parameter} & \textbf{Specification} \\ \hline \hline
            CPU type & Intel(R) Core(TM) i7-7700 CPU \\
            CPU core count & 4 \\
            Frequency & 3.6 GHz \\ \hline
            L1 cache per core & 32 KiB + 32 KiB (I + D) \\
            L2 cache & 1 MiB \\
            L3 cache & 8 MiB \\ \hline
            DRAM manufacturers & SK Hynix, Samsung \\
            \multicolumn{1}{l}{DRAM technology} & DDR4 \\
            DRAM size & 8 GB \\ \hline
            \end{tabular}%
    \end{table}

    \begin{figure}[!b]
        \centering
        \mbox{
        \subfigure[Run 1]
            {
            \includegraphics[width=0.08\textwidth]{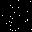}
            }
        \subfigure[Run 2]
            {
            \hspace{-2ex}
            \includegraphics[width=0.08\textwidth]{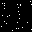}
            }
        \subfigure[Run 3]
            {
            \hspace{-2ex}
            \includegraphics[width=0.08\textwidth]{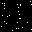}
            }
        \subfigure[Run 4]
            {
            \hspace{-2ex}
            \includegraphics[width=0.08\textwidth]{images/real_run_3.png}
            }
        \subfigure[Superimposed Image]
            {
            \hspace{-2ex}
            \includegraphics[width=0.08\textwidth]{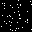}
            \label{fig:map}
            }
        }
        
        \caption{Observed bitflips on a single row on a width 8 DRAM DIMM (row 1278). There are 1024 columns
        per row. Each of these images presents a 32x32 shaped 2D image of a
        single row. A white dot represents a flipped column.}
        \label{fig:rh-real-map}
    \end{figure}

    \subsection{Hardware Characterization Experiments}
        To ground our architectural probability models in physical ground truth, we conducted extensive hardware characterization experiments across multiple commercially available DDR4 DIMMs.
        The goal of these experiments is to quantify the exact spatial and temporal probability distributions governing RowHammer failures under controlled environments.

        We deployed the Blacksmith~\cite{blacksmith,blacksmith-code} profiling infrastructure on a desktop detailed in Table~\ref{tab:kaby-lake}.
        We executed double-sided RowHammer profiling by targeting a specific victim row sandwiched between two continuously activated aggressor rows.
        Each profiling loop executed $5\times10^6$ \texttt{ACTs}, and the routine was repeated over 1,000 distinct iterations to observe long-term bitflip patterns, cumulative error saturation, and localized vulnerability profiles.

    \subsection{Empirical Observations}
        Our physical characterization yielded three core insights regarding the nature of modern DRAM vulnerabilities, which serve as the direct mathematical foundations for the \papertitle architectural abstraction layers:

        \begin{framed}
            \noindent\textbf{Observation 1 (Spatial Localization):} RowHammer-induced bitflips are intensely localized within the memory array.
            While specific regions of a given DRAM row are highly vulnerable to RowHammer, vast adjacent structures remain entirely resilient.
        \end{framed}

        This behavior (Figure~\ref{fig:rh-real-map}) indicates that local manufacturing process variations dictate cell weakness.
        Rather than hard-coding these patterns for a single device, \papertitle resolves \textit{C1} by providing two generalized modeling paths: an ingested, file-based physical \texttt{device map}, or a statistical process variation engine based on VARIUS~\cite{varius} that generates multivariate normal distributions across the simulated memory coordinates to flag distinct \textit{Weak Cells} (\texttt{WC}).

        \begin{figure}[!b]
            \centering
            \includegraphics[width=0.49\textwidth]{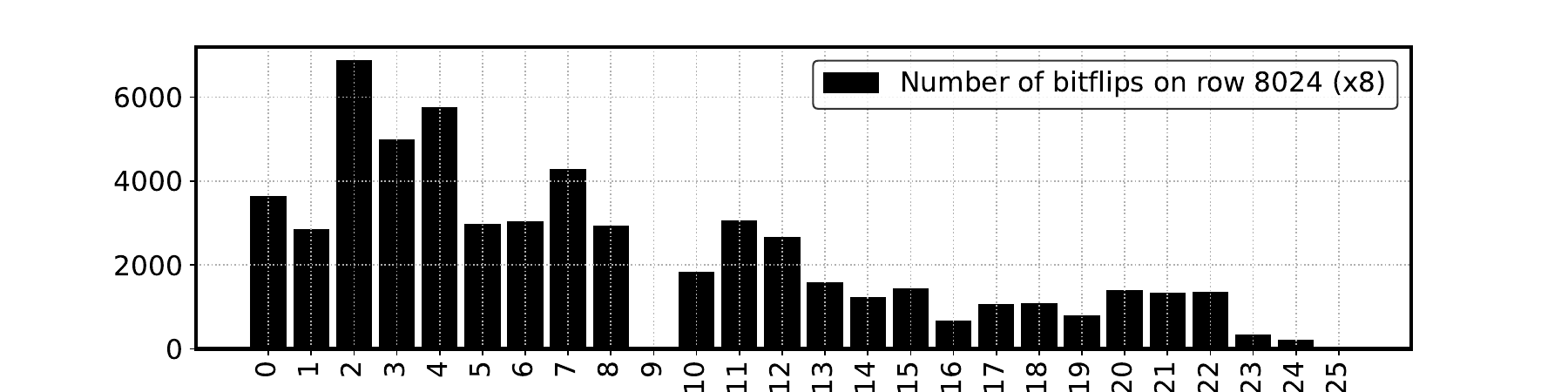}
            \includegraphics[width=0.49\textwidth]{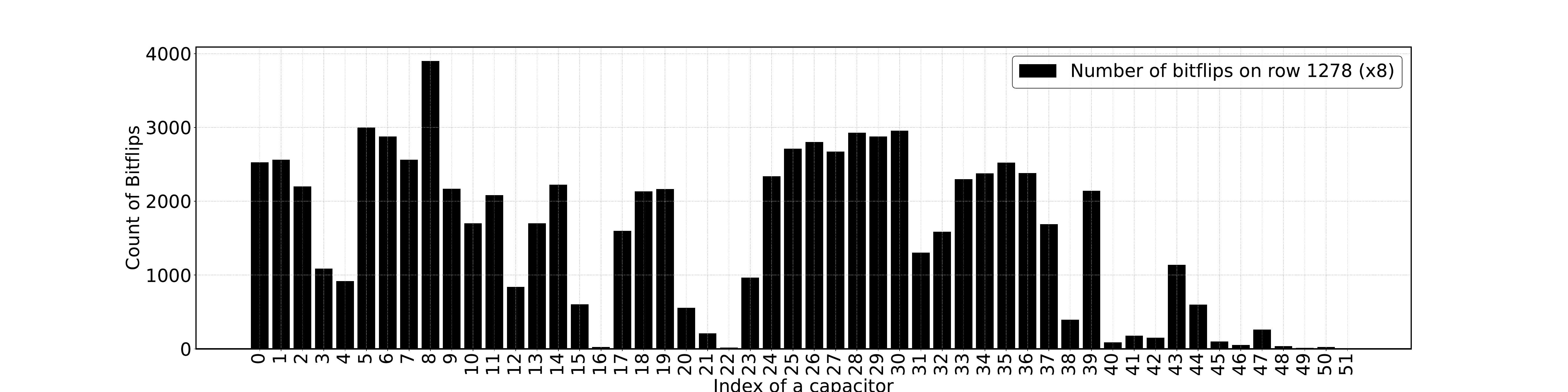}
            \caption{Histogram of bitflip counts on row number 8024 (top) and
            1278 (bottom) across 1000 runs.}
            \label{fig:rh-histogram}
        \end{figure} 

        \begin{framed}
            \noindent\textbf{Observation 2 (Temporal Saturation):} The unique bitflip count across a DRAM array saturates over time.
            Cells that fail during an initial activation cycle demonstrate a bounded, distinct probability $p$ of failing again across subsequent cycles.
        \end{framed}

        The empirical histograms in Figure~\ref{fig:rh-histogram} shows this saturation behavior.
        This establishes that the pool of vulnerable cells is structurally finite.
        Once a simulated cell is identified as a \texttt{WC}, \papertitle evaluates a uniform probability distribution at runtime to determine if an active activation stresses the cell into a flippable state (\texttt{F\_WC}).
        We measured this baseline bitflip probability between $\frac{1}{5 \times 10^{10}}$ and $\frac{1}{5 \times 10^{8}}$ per activation, which we expose as a user-tunable parameter.

        \begin{framed}
            \noindent\textbf{Observation 3 (Attack Proximity Non-Linearity):} The probability of inducing a data failure scales non-linearly by multiple orders of magnitude when executing coordinated multi-sided RowHammer patterns relative to single-sided access sequences.
        \end{framed}

        Our physical tracking demonstrated that executing a double-sided RowHammer sequence amplifies the bitflip probability by a factor of $2.5 \times 10^3$ compared to a single-sided baseline hitting the identical weak cell cluster.
        To implement this insight, \papertitle leverages dynamic tracking counters inside the memory interface, tracking the non-linear threat scaling of single-sided, double-sided, and generalized multi-row patterns.

        \begin{figure}[t]
            \centering
            \includegraphics[width=0.49\textwidth]{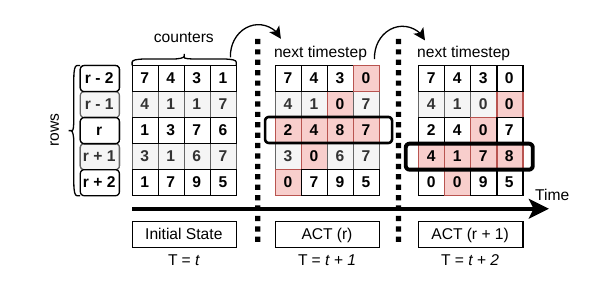}
            \caption{\papertitle~maintains structural counter-pairs per DRAM row to 
            to dynamically track multi-row RowHammer access patterns. All counters are reset at each \texttt{tREFW} boundary.
                This figure shows 3 states of these counters: initial, after reading row \textit{r} and after reading row \textit{r + 1}.}
            \label{fig:hammersim_counters}
        \end{figure}

    \subsection{Architectural Implementation}
        To incorporate these empirical insights into a unified simulation framework, \papertitle introduces structural and systemic modifications to the \gem~DRAM interface, implementing both online and offline execution modalities.
        In Appendix~\ref{app:using_hammersim}, we provide details on navigating the \papertitle source-code.

    \subsubsection{Counter-Pairs and Adjacency Tracking}
        To resolve \textit{C2} and capture complex spatial interactions without exponential simulation overheads, \papertitle maintains optimized tracking structures within the memory controller.
        As illustrated in Figure~\ref{fig:hammersim_counters}, each row tracks its immediate neighborhood using counter-pairs.
        For standard vulnerabilities, this mechanism monitors single and double-sided access rates. 
    
        To support advanced vulnerabilities like Half-Double exploits~\cite{halfdouble}, we expand the tracking radius to monitor four dedicated row activation counters ($r\pm1, r\pm2$) relative to the target row $r$.
        To optimize performance, these structures evaluate fault triggers during standard refresh intervals (\texttt{tREFW}), clearing all tracking metadata at each window boundary.

    \subsubsection{Online Mode: Runtime Fault Injection and Functional ECC}
        In \textit{online mode}, \papertitle couples its hardware-informed probability distributions directly into \gem's operational memory execution path.
        Every memory request dynamically triggers fault evaluation checks based on current row activation states, injecting live, bit-level data corruption into the simulator's active memory arrays. 

        Because physical RowHammer bitflips occur silently within the DRAM capacitors without throwing hardware interrupts, they act as transient soft errors.
        \papertitle replicates this behavior: while all injected errors are logged to standard output (\texttt{stdout}) for diagnostic tracking, the simulated application and operating system remain unaware of the corruption until they actively address or validate the corrupted memory address.
        This allows for authentic evaluation of silent data corruption across OS-level page schedulers, security kernel modules, and user space applications.

        To support the exploration of cross-layer resilience, we implement a functional Error Correcting Code (ECC) abstraction layer within \gem's memory subsystem.
        ECC-enabled DIMMs~\cite{ddr_ecc} are a popular option for server-class systems~\cite{ecc_rowhammer}.
        RowHammer on ECC-enabled DIMMs is a popular topic of research for both attacks~\cite{ecc_rowhammer} and defenses~\cite{polymorphic-ecc,rampart,rethinking_ecc}.

        We implement SECDED algorithm from the reverse-engineered paper by Cojocar \textit{et al.}~\cite{ecc_rowhammer}.
        \papertitle pads every word with parity blocks calculated via a user-defined text evaluation matrix (\texttt{pMatrix}).
        At runtime, when a corrupted row is accessed, the ECC module intercepts the request, evaluates the parity matrix, and dynamically updates native \gem~telemetry statistics (\texttt{stats.txt}) to record correctable single-bit faults or fatal, uncorrectable multi-bit errors.

    \subsubsection{Offline Mode: Decoupled Trace Analysis}
        To address \textit{C3} and enable rapid design space exploration without the penalty of repetitive, long-running full-system simulations, \papertitle introduces an optimized \textit{offline mode}.
        When active, the memory controller logs row-level activation (\texttt{ACT}) allocations to a structured trace file at every \texttt{tREFW} boundary. 

        An external analytical tool subsequently processes the RowHammer trace that overlays the workload's spatial activation footprint against alternative process variation distributions, device maps, or mitigation thresholds.
        This decoupling allows hardware designers and security analysts to evaluate how an identical application execution profile interacts with dozens of distinct DRAM manufacturing variations instantly, eliminating the need to re-run the underlying architectural simulation.

        In Appendix~\ref{app:params}, we list all the available \papertitle simulation parameters introduced.

\section{Discussion and Limitations}
    \subsection{RowHammer Research Enabled}
        \papertitle~allows researchers to investigate RowHammer attacks, defenses, and, also potentially study applications of RowHammer.
        In this section, we discuss several use cases that \papertitle enables.

        \subsubsection{RowHammer Probability Distribution Study}
            \papertitle allows researchers to investigate different probability distributions of causing a bitflip.
            We incorporate known probabilities from prior works~\cite{centauri,para,trrespass,rowhammer-puf,halfdouble,half-double-paper} and our experimental observations.
            However, there is no mathematical evidence that these are all the distributions that define RowHammer.
            The probability of a bitflip of a strong DRAM cell, for example, is still missing from our evaluation.
            \papertitle can be extended in the future to incorporate such distributions, when discovered, and studied in a full-system environment.

        \subsubsection{RowHammer Attack}
            \papertitle~allows users to uncover sophisticated memory access patterns that cause RowHammer attacks.
            Creating a RowHammer attack as a memory trace is straightforward by using gem5 traffic generators.
            Such traffic generators produce deterministic access patterns for a given device map.
            Various RowHammer attacks like Juggernaut attack pattern~\cite{row-swap-extension}, double RowHammer attack~\cite{cratar}, \textit{etc.}, can be created using \papertitle.
            These maps can be refined by researchers or vendors using process variation data.

            In contrast, implementing RowHammer at the system level is far more complex.
            Prior attacks~\cite{windows_rowhammer,seaborn2015exploiting,anvil,trrespass,blacksmith,sledgehammer,smash,rowhammerjs} show that bypassing OS abstractions is challenging.
            While evicting caches suffices for single-sided RowHammer on x86 systems, obtaining contiguous rows on the same bank in multi-channel, multi-rank servers remains difficult~\cite{centauri}.
            Furthermore, physical address mapping strongly influences RowHammer bitflip behavior~\cite{drama,trrespass}.

            Further, \papertitle can model GPU-based RowHammer attacks~\cite{drammer,gpuhammer,gpu-hammer1,gpu-hammer2} using the GPU model in gem5~\cite{gem5-gpu,gem5-amd-gpu}.
            


        \subsubsection{RowHammer Mitigation}
            \papertitle~enables users to prototype and evaluate hardware-based RowHammer mitigations at the system level.
            It supports two key components: (a) detection and (b) inhibition, modeled with timing correctness.
            To demonstrate this flexibility, HammerSim replicates simplified TRR mechanisms based on reverse-engineered designs~\cite{utrr, blacksmith} from major DRAM vendors, including counter-based and probabilistic refresh strategies.
            These implementations show how researchers can extend \papertitle's data structures to explore new mitigation techniques without breaking DRAM timing constraints.

            TRR refreshes are sent out when the DRAM DIMM is locked (for tRFC time) during refreshes (via REFI instructions).
            Most research-based RowHammer mitigations, on the other hand, explicitly send out refreshes as \texttt{ACTIVATE} to the victim rows, violating DRAM timings.
            DRAM devices adhere to strict timing requirements~\cite{ddr3, ddr4, ddr4-samsung,ddr5, gddr5, gddr6}.
            TRR~\cite{ddr4-samsung, trrespass, blacksmith} sends targeted \text{REF} commands~\cite{trrespass} (or simply \texttt{ACT} commands) to possible victim rows.
            
            \papertitle~allows simple extensions to the existing data structures to implement a RowHammer mitigation techniques like~\cite{para,twice,graphene,crow,blockhammer,cbt}.
            Further, \papertitle allows tweaking such mechanisms to adhere to the timing contraints set by existing TRR mechanisms via the Inhibitor module.
            To show that, we implement some of the classic and \textit{state-of-the-art} RowHammer mitigations like PARA~\cite{para} and TWICE~\cite{twice}. 

        \subsubsection{Error Correcting Code}
            Another direction of RowHammer mitigation is ECC~\cite{ecc_rowhammer,polymorphic-ecc,rethinking_ecc,rampart}.
            ECC-enabled DIMMs are commonly available in the market today.
            DDR5 DIMMs are shipped out with an on-chip proprietary ECC implementation~\cite{zenhammer,ddr5-ecc-rowhammer}.
            
            Qureshi \textit{et al.} suggested investigating RowHammer defenses via newer ECC algorithms~\cite{rethinking_ecc}.
            Online \papertitle~enables functional ECC modeling to study RowHammer defenses in a full-system context.
            It supports selective error correction during simulation, allowing researchers to evaluate ECC schemes under realistic workloads.
            For demonstration, we implement a SECDED algorithm (1-bit correction, 2-bit detection) using Cojocar \textit{et al.} reverse engineered ECC algorithm~\cite{ecc_rowhammer} on a DDR3 DIMM within \papertitle.

            ECC bits are computed from a user-supplied \texttt{pMatrix} and applied on-demand to minimize simulation overhead.
            This capability allows exploration of new ECC strategies and their effectiveness against RowHammer-induced corruption.
            


        
        \subsubsection{RowHammer Applications}
            \papertitle~enables researchers to explore RowHammer-based applications such as PUFs~\cite{rowhammer-puf,rowhammer_attack_puf} and DRAM fingerprinting~\cite{centauri,fphammer} by modeling cell-level variation maps in simulation.
            Unlike prior simulators that assume uniform vulnerability, HammerSim uses device maps derived from real DIMMs or statistical models to capture per-cell variation.
            This capability also supports evaluating defenses like row-swapping~\cite{row-swap,row-swap-extension,rhlm} by identifying strong and weak cells, reducing overhead while preserving security.
            Further, in Section~\ref{sec:online-eval}, we show how \papertitle can be used to study RowHammer-based soft-errors~\cite{softhammer}.

    \begin{figure*}[t]
        \centering
        \mbox {
            \subfigure[JS divergence heatmap of 9 different rows on dimm0] {
                \label{fig:js_dimm0}
                \includegraphics[width=0.3\textwidth]{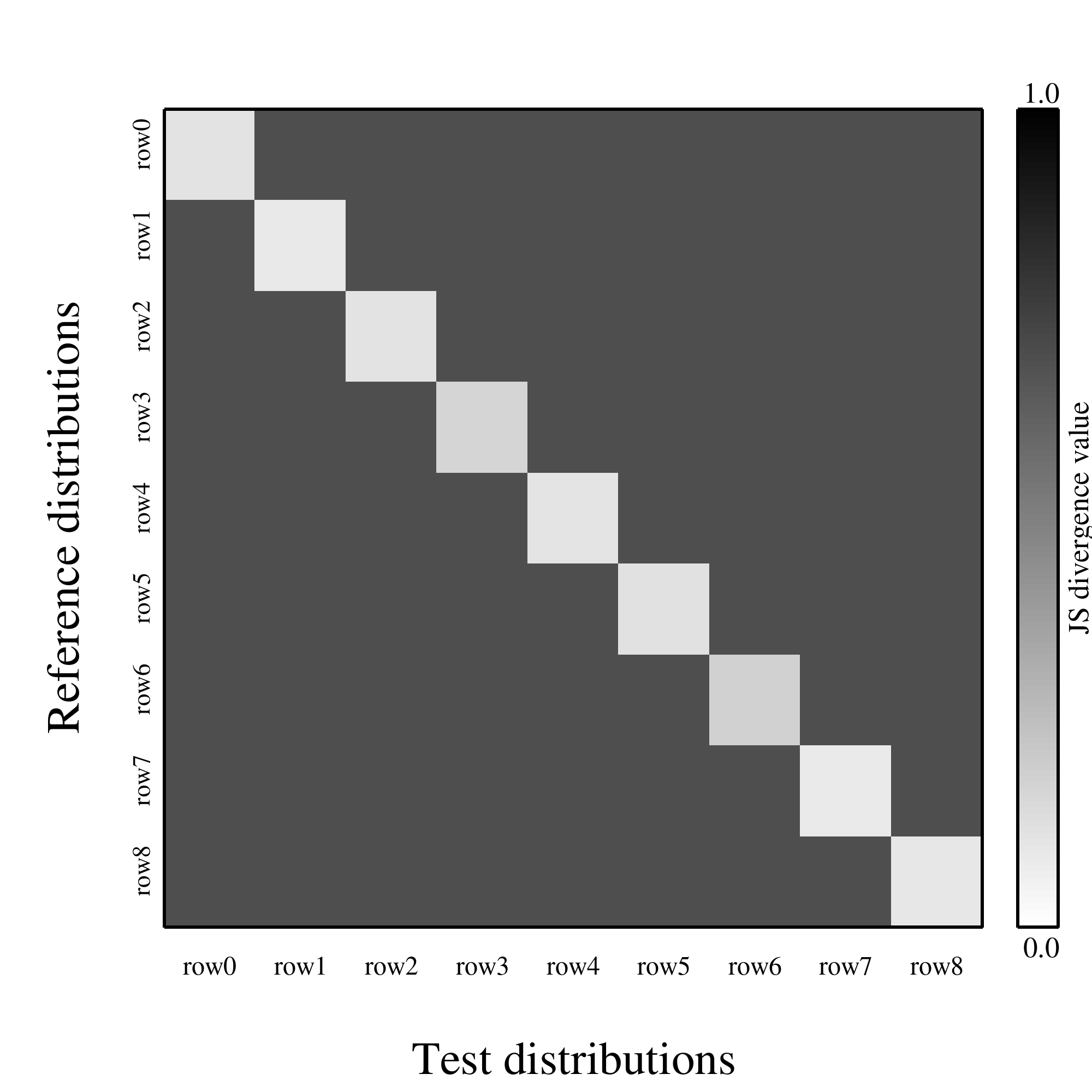}
            }
            \subfigure[JS divergence heatmap of 9 different rows on dimm1] {
                \label{fig:js_dimm1}
                \hspace{-2ex}
                \includegraphics[width=0.3\textwidth]{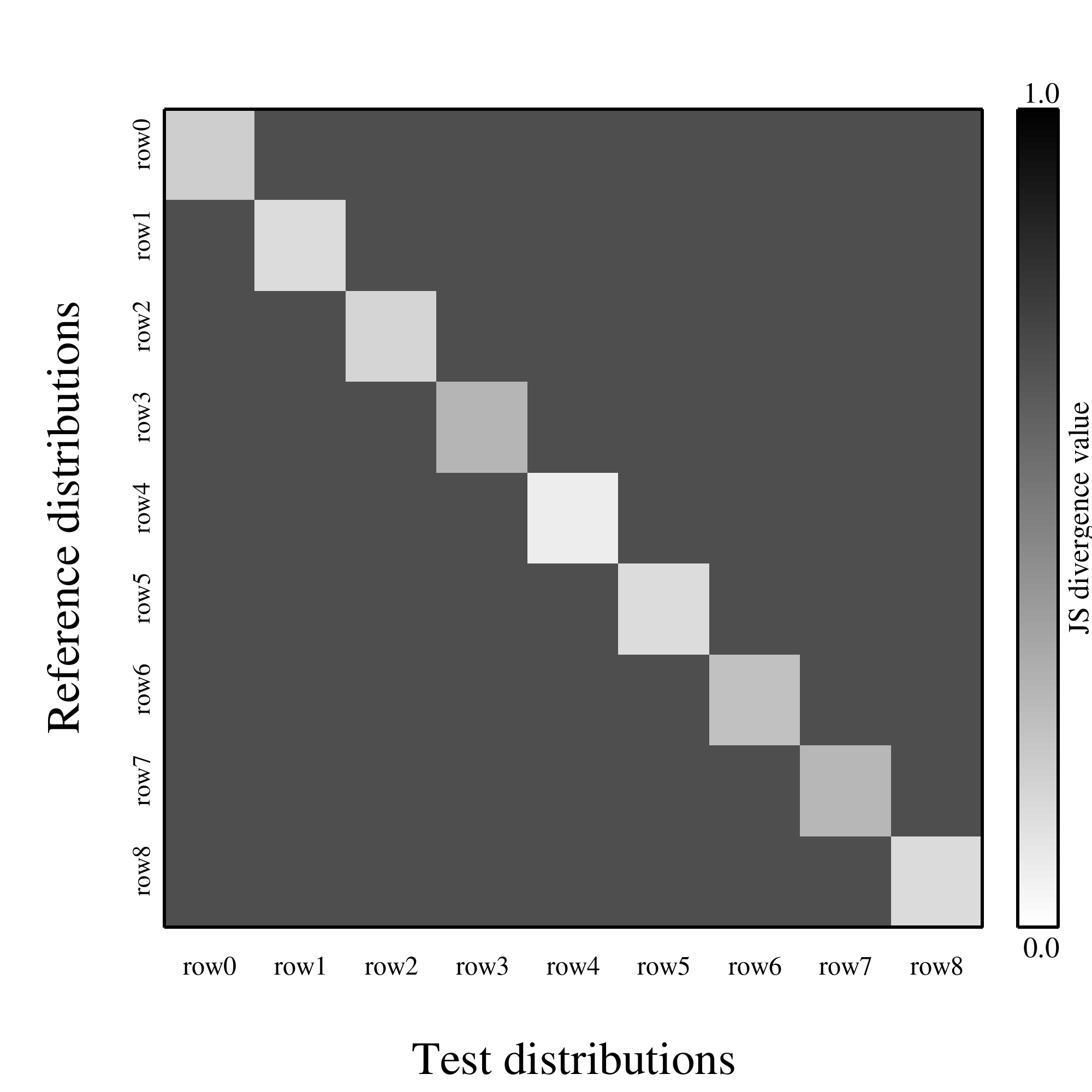}
            }
            \subfigure[JS divergence heatmap of 9 different DIMMs from the same DRAM manufacturer hammering row 7292] {
                \label{fig:js_accross_banks-b}
                \hspace{-2ex}
                \includegraphics[width=0.3\textwidth]{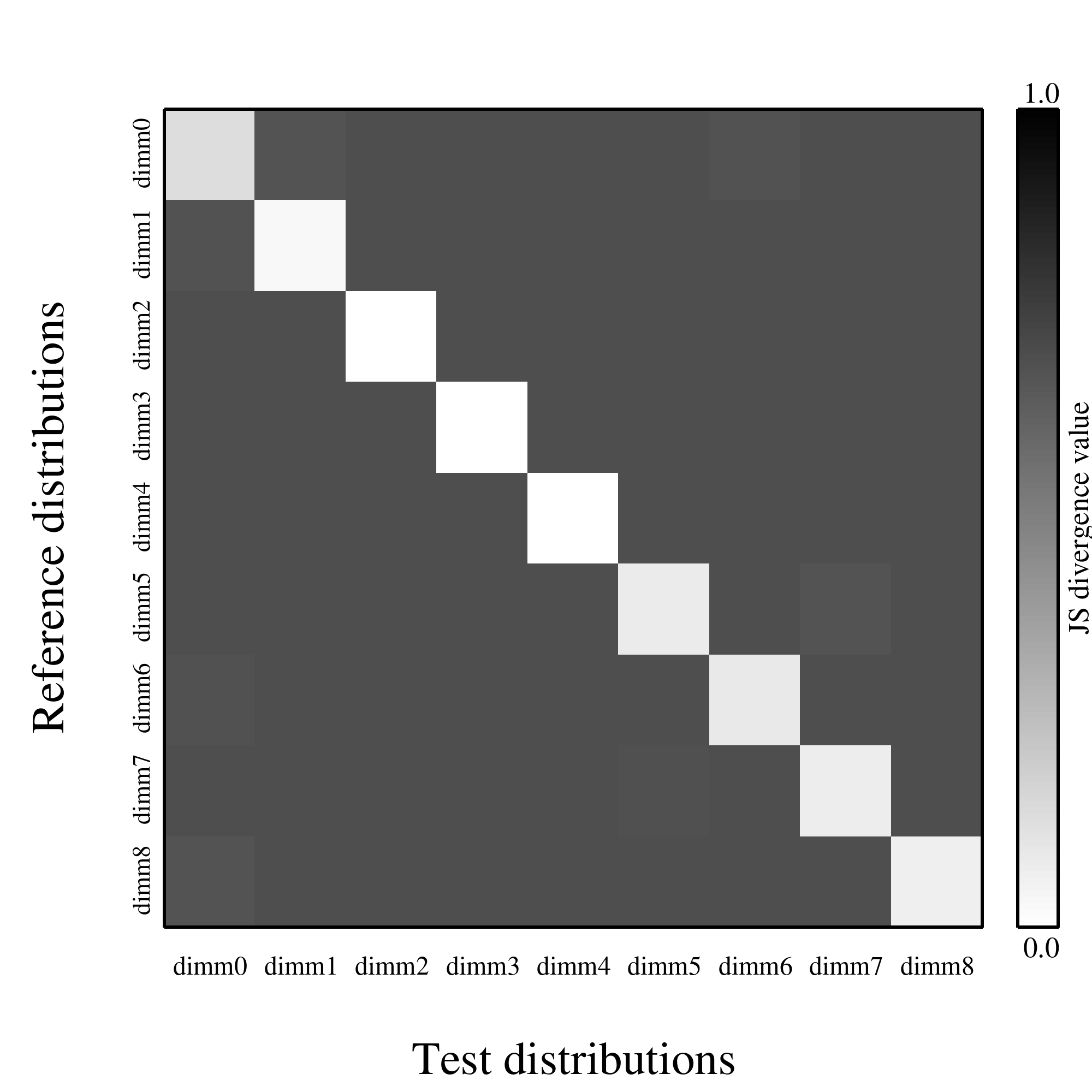}
            }
            }
            \caption{JS divergence heatmaps across different victim rows of the same DIMM and across multiple DIMMs on real hardware and \papertitle.}
            \label{fig:js_accross_rows}
        \end{figure*}
    \subsection{Limitations}
        \label{sec:limitations}
        \papertitle~introduces invasive changes to \gem's memory model by embedding RowHammer tracking and mitigation logic directly in the memory controller interface, rather than using separate SimObjects.
        This design mirrors real DRAM behavior, where TRR operates on-chip.
        However, this approach requires periodic updates and rebasing as gem5 evolves.
        Making changes to the source is intuitive as our RowHammer trackers get triggered when \texttt{activateBank()} is called and the inhibitor is called at \texttt{processRefreshEvent()}, similar to the on-chip logic of a typical DIMM~\cite{trrespass}.

        A key limitation is the abstraction gap: simulators~\cite{dramsim2,dramsim3,ramulator} model DRAM at the logical level (banks and cache lines), whereas physical DIMMs activate entire rows spanning multiple chips.
        \papertitle~currently applies bitflips within the same logical bank as the aggressor, while ensuring the correctness of the victim rows.
        This means that the data is tracked at the cache line level using simulator-dependent data objects.

        Another limitation is reliance on statistically generated variation maps.
        While multivariate models approximate process variation and are popular~\cite{process}, they may deviate from real DIMM characteristics.
        To address this, \papertitle~supports user-supplied device maps derived from hardware experiments (e.g., Blacksmith~\cite{blacksmith}, TRRespass~\cite{trrespass}).

        ECC is functionally implemented in \papertitle.
        Unlike a real ECC DIMM that stores the checksum bits on the DRAM array, we compute the checksum on the fly to model functional correctness.
        This allows newer ECC algorithms to be evaluated online during simulation.
        However, checksum data corruption and timing correctness of the memory are not modeled.

        Finally, \papertitle~does not model true and anti cells~\cite{para}, as these mappings are proprietary.
        A charged DRAM capacitor does not necessarily mean a value of one~\cite{para,blacksmith,trrespass}.
        Extending \papertitle~to enable such mappings is trivial and can be implemented the same way as the device map.
        This does not affect the RowHammer probabilities of a bitflip.

        


\section{Evaluation}
    \label{sec:results}

    \papertitle is implemented in gem5~\cite{lowe2020gem5}, a full-system cycle-level simulator.
    We model the probability distributions of RowHammer (Section~\ref{sec:method}) within gem5's memory interface.


    \papertitle~allows users to perform several kinds of experiments. 
    We narrow down the search space and perform three case studies to evaluate \papertitle.
    We fix our experimental system as a commodity-class system in most of our experiments.
    The parameters of the system are described in Table~\ref{tab:params}.

    We validate our simulation model against real hardware in Section~\ref{sec:case-1}.
    In Section~\ref{sec:case-2}, we show how a full system RowHammer application can be studied in \papertitle~and how ECC can functionally correct bitflips.
    Finally, we also demonstrate how benign applications are affected by RowHammer as the RowHammer threshold is decreasing with each DRAM generation.

    
    
\begin{table}[t]
    \centering
        \caption{Simulated System Parameters}
        \label{tab:params}
    \begin{tabular}{cc}
    \hline
    \textbf{Parameter} & \textbf{Specification} \\ \hline \hline
    CPU type & TimingSimpleCPU \\
    CPU core count & 1/2/8 \\ \hline
    L1 cache per core & 32 KiB + 32 KiB (I + D) \\
    L2 cache & 256 KiB \\ \hline
    DRAM technology & DDR3/DDR4 \\
    DRAM size & 1/2 GB \\ \hline
    \end{tabular}%
\end{table}

    \subsection{Case Study I: Validation}
        \label{sec:case-1}
        We want \papertitle~to model existing DIMMs inside the simulator.
        This implies that we also need to model TRRs installed in the DIMMs.
        Towards this, we have implemented RNG-based TRR mitigation techniques in \papertitle, conceptually similar to U-TRR~\cite{utrr}.
        We use Blacksmith~\cite{blacksmith}'s fuzzer to find access patterns on the same DIMMs.
        Bitflips are generated per bank.
        We record the aggressor traffic.
        We encode the bitflip maps generated from the hardware and plug in our simulation as the variation map.
        
        Our goal in this case study is to model a similar bitflip map in simulation.
        We split the test cases into within rows of a single DIMM and within banks of multiple DIMMs.
        Blacksmith's fuzzer bypasses TRR with a specific memory access pattern.
        We use \gem's traffic generator to generate synthetic traffic for the aggressor rows, recreating the attack.
        \gem's traffic generator can send a large amount of traffic at the minimum interval of 1 ps.
        We adjust our traffic pattern to mimic real systems with caches and tune the RowHammer probabilities to account for such high traffic rates. 
        Note that the device address mapping scheme, \textit{i.e.}, physical to the DRAM device mapping, is different in the experimental system and the simulated system.
        We make sure that the reference and the test systems have compatible mapping.

        We compare the weak columns of hardware-generated bitflips with the simulated ones for single rows.
        We use JS Divergence~\cite{js-divergence} as the similarity metric to compare bitflip maps.
        JS divergence is a method of measuring the similarity between two probability distributions.
        A lower JS divergence value means the two distributions are more similar.
        We present our results as a heatmap of JS divergence values with references collected from the hardware and tested using \papertitle.
        We select 9 rows from two different DIMMs from the same DRAM manufacturer and show our findings in Figure~\ref{fig:js_dimm0} and Figure~\ref{fig:js_dimm1}.
        We see smaller JS divergence values for the same reference and test case, suggesting a closer to hardware model of RowHammer in \papertitle.


        Next, we repeat the same experiment across multiple DIMMs.
        We select 9 different DIMMs from the same manufacturer and model the RowHammer device map in \papertitle. 
        We plot the heatmap of JS divergence values in Figure~\ref{fig:js_accross_banks-b}, which shows a close resemblance to the real hardware.




        \begin{framed}
            \noindent\textbf{Takeaway 1}: 
            \papertitle~allows localized RowHammer bitflip simulation via the device map.
            Weak cells per row can be modeled with high fidelity in \papertitle, up to the capacitor level, given that TRR does not interfere.
            While a correct reverse-engineered TRR implementation is not the goal of this paper, we show how \papertitle~can be used to model the entire DIMM in simulation for higher correctness. 
        \end{framed}


    \subsection{Case Study II: Full-System Applications and Data Corruption}
        \label{sec:case-2}
        We show the execution of Google's \texttt{rowhammer-test}~\cite{seaborn2015exploiting} inside the simulator in a full-system setting.
        The OS abstractions are bypassed by rowhammer-test, as it can perform both single-sided and double-sided RowHammer attacks.
        Our objective is to test whether we can use rowhammer-test as a standard RowHammer benchmark workload.

        \begin{figure}[!b]
            \centering
            \includegraphics[width=0.48\textwidth]{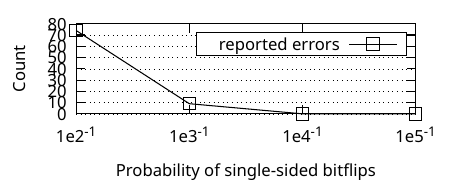}
            \caption{Reported bit errors by the simulator.}
            \label{fig:rowhammer-test-stats}
        \end{figure} 

\begin{figure}[h]
            \centering
            \includegraphics[width=0.48\textwidth]{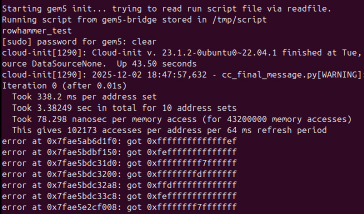}
            \caption{\texttt{stdout} of \texttt{rowhammer-test} when the \textit{single-sided} RowHammer attack was performed online using \papertitle.}
            \label{fig:rowhammer-test-stdout}
        \end{figure}

        We simulated \texttt{rowhammer-test} inside \papertitle~using a reproducible gem5-resource~\cite{gem5-resource} disk-image.
        We hammered the memory using the single-sided RowHammer attacks from \texttt{rowhammer-test} on a system specified in Table~\ref{tab:params}.
        We ran a simulation of one iteration of \texttt{rowhammer-test}.
        We show the bitflip count in Figure~\ref{fig:rowhammer-test-stats} by varying the probability of a bitflip.
        \texttt{rowhammer-test}'s output is printed on \texttt{stdout} with the number of bitflips and the change in data, as shown in Figure~\ref{fig:rowhammer-test-stdout}.

        As mentioned in Section~\ref{sec:limitations}, we do not model true and anti-DRAM cells.
        For a bitflip, we flip the contents of the capacitor regardless of a charged or an unchanged capacitor.
        \texttt{rowhammer-test} writes \texttt{0xFF} as the data pattern and compares it for bitflips later.
        Bitflips are XOR on the exact capacitor location.

        \begin{framed}
            \noindent\textbf{Takeaway 2}: 
            \papertitle~enables simulated data corruption shown using \texttt{rowhammer-test}.
            This allows system-level RowHammer attacks, defenses, and applications to be studied in simulation. 
        \end{framed}




    \begin{table}[!b]
        \centering
            \caption{SECDED ECC statistics of correcting RowHammer-induced bitflips as reported by \papertitle. Bitflips are generated randomly based on the user-supplied probability.}
            \label{tab:ecc}
        \resizebox{0.48\textwidth}{!}{%
\begin{tabular}{c|cccc|c}
\hline
\textbf{Probability} & \textbf{Bitflips} & \textbf{\begin{tabular}[c]{@{}c@{}}Single\\ Errors\\ Corrected\end{tabular}} & \textbf{\begin{tabular}[c]{@{}c@{}}Double\\ Errors\\ Detected\end{tabular}} & \textbf{\begin{tabular}[c]{@{}c@{}}Uncorrectable\\ Errors\end{tabular}} & \textbf{Notes} \\ \hline \hline
$1e1^{-1}$ & 237 & 29 & 2 & 17 & Kernel panic \\
$1e2^{-1}$ & 115 & 3 & 0 & 1 & None \\
$1e3^{-1}$ & 95 & 1 & 0 & 0 & None \\
$1e4^{-1}$ & 0 & 0 & 0 & 0 & None  \\ \hline
\end{tabular}
        }
    \end{table}
        We extend the previous case study to test ECC within the simulator.
        We functionally model ECC, that does not account for timing changes in the DRAM.
        At a bitflip, we store the original data for ECC parity bit computation.
        When victim rows are later accessed, we call the ECC algorithm to keep the wall clock time of the simulation low.
        Whenever RowHammer triggers a bitflip, we compute the \texttt{gMatrix} based on an 8-byte word and the user-supplied \texttt{pMatrix}.

        For single-bit errors on an 8-byte word, we use the ECC bits to correct the data when the victim row is accessed.
        \papertitle~reports the bitflips as an output statistic.
        We keep the simulation parameters the same as before.
        As we increase the probability of a bitflip, we see multiple bitflips on the same row (Table~\ref{tab:ecc}).
        The SECDED ECC corrects up to one bit flip and detects up to 2 bit flips~\cite{ecc_rowhammer}.
\\ \\

        \begin{framed}
            \noindent\textbf{Takeaway 3}: 
            Data corruption in \papertitle naturally allows the simulator to investigate newer and stronger error correction codes (ECC).
        \end{framed}

    \subsection{Case Study III: Benign Applications}

         \subsubsection{Offline Analysis}

            To demonstrate \papertitle's offline analysis, we dump the memory access traces of two popular benchmark suites: NAS-Parallel benchmarks~\cite{bailey1991parallel_bk} and GAPBS~\cite{beamer2015gap}.
            In our experiment, we did not see any RowHammer access patterns in these benign applications with the standard bitflip probability parameters.
            With shrinking process size, the RowHammer threshold is also decreasing~\cite{trrespass,para}.
            Therefore, we perform an offline estimation of bitflips on these memory traces.

\begin{figure}[t]
                 \centering
                 \mbox{
                     \subfigure[Gapbs] {
                     \label{fig:result-5}
                     \includegraphics[width=0.45\textwidth]{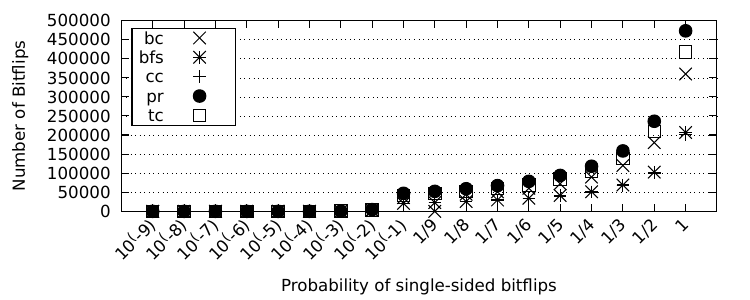}
                     }
                 }
                 \mbox{
                     \subfigure[NAS parallel benchmarks] {
                         \label{fig:result-6}
                         \includegraphics[width=0.45\textwidth]{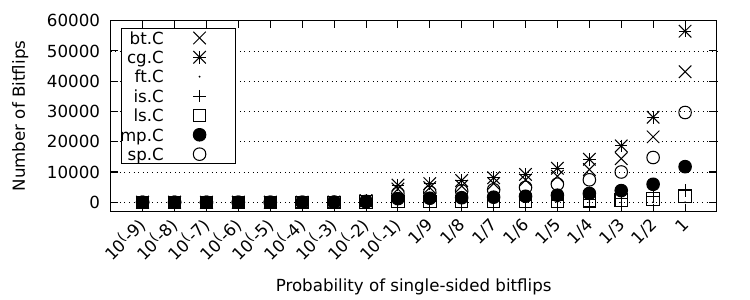}
                     }
                 }
                 \caption{Offline estimation of bitflips by varying the probability of a bitflip for a DDR4 DRAM device.
                         Note that the x-axis in both of these graphs are discontinous as it progresses in factors of 10 until (8, 0), and then it progresses in steps of 0.1.}
             \end{figure}

            \papertitle~allows us to tune probabilities of bitflips to correct over/under estimations via the post-processing tool.
            We show this by varying the probability of observing a single-sided RowHammer-induced bitflip from $10^{-9}$ to $1$.
            The results obtained are depicted in Figure~\ref{fig:result-5} and
            Figure~\ref{fig:result-6}.

            The memory access pattern for NPB and GAPBS never had double-sided RowHammer patterns.
            First, we experiment assuming a uniform distribution of observing bitflips and did not rely on the device map.
            We see that there are 20041.16 instances on average across the two benchmark suites, where rows cross the DDR4 threshold of 45K ACTs.

            We repeat the same evaluation using a statistically generated device map for RowHammer bitflip distribution.
            The idea is to ignore \textit{strong} DRAM rows, which are resistant to RowHammer attacks.
            We see that on average, there is a reduction of 90.1\% probability of observing a bitflip compared to the previous set of results.
            Our observation inside a simulator is consistent with bitflip patterns observed in prior works~\cite{centauri, drammer, rowhammer-puf, blacksmith_recreate_paper,fphammer}, where authors of the respective papers report localized RowHammer-induced bitflips.

            \begin{framed}
                \noindent\textbf{Takeaway 4}: This experiment shows that \papertitle~can be used to investigate RowHammer access patterns at a finer granularity while enabling users to estimate RowHammer bitflip probability.
            \end{framed}

        \subsubsection{Online Analysis and Data Corruption Study}
             \label{sec:online-eval}
            In our experiments, we did not see any instances of RowHammer bitflips with the default RowHammer parameters of a DDR4 DRAM DIMM.
            We repeated the experiment with a lower RowHammer threshold (1000) and a higher RowHammer probability ($1e1^{-1}$).
            However, we are still interested to observe soft errors in benign applications.
            Towards this, we repurpose \papertitle's infrastructure to study RowHammer-induced bitflips in the neighborhood of the aggressor row with a very low RowHammer threshold in benign applications.
            The goal is to observe silent error propagations on benign applications when the RowHammer threshold becomes dangerously low for future shrinking process.

            We selected smaller workload sizes from our two selected benchmark suites (NPB class \textit{A} and GAPBS size \textit{-g 20}).
            The goal is to finish at least one iteration of ROI and validate the results on a system without caches to maximize memory accesses.
            We set up the experiment with data corruption enabled, set the RowHammer threshold (1), and the probability to study soft errors ($1e^{-1}$).

            We model a system with a single CPU core without caches and 1 GiB DDR4 DRAM memory.
            For the baseline, we simulate an ideal DDR4 DRAM device without the possibility of any bitflips to measure the host system's execution time, total memory consumption, and output correctness.
            We repeat this experiment by replacing the DRAM with RowHammer-affected ones and simulate the same workload with and without data corruption.

            Table~\ref{tab:benign-app-live} shows the simulation statistics of \papertitle and baseline gem5.
            Overall, we saw \papertitle takes 10.9\% and 29.2\% additional wall clock time on average to simulate RowHammer DIMMs compared to a baseline unmodified gem5 for NPB and GAPBS, respectively.
            To maintain the additional data structures to simulate RowHammer, \papertitle consumes 50.6\% and 65\% additional host memory for NPB and GAPBS, respectively.

            Table~\ref{tab:benign-app-live} also shows the outcome of the experiments, \textit{i.e.}, errors induced via RowHammer.
            While every workload had bitflips (column 6), not all workloads failed to reach the end of ROI.
            This means that there were silent errors that propagated to the end of the program.
            We did validate the results at the end of the ROI.
            We saw that workloads like $mg$ and $ft$ successfully validated their results, which means that RowHammer bitflips did not affect the outcome of the workload.
            This has been shown in prior research as well~\cite{softhammer}.
            We plan on investigating this aspect of fault-tolerant workload behavior in the future.

            \begin{framed}
                \noindent\textbf{Takeaway 5}: While \papertitle is 11.9\% slower than gem5 overall, it allows the study of soft errors on full-system applications.
            \end{framed}

\begin{table}[]
\centering
\resizebox{0.49\textwidth}{!}{%
\begin{tabular}{c|cc|cc|cc}
\hline
\multirow{2}{*}{\textbf{Benchmark}} & \multicolumn{2}{c|}{\textbf{\begin{tabular}[c]{@{}c@{}}Host\\ Runtime\\ (s)\end{tabular}}} & \multicolumn{2}{c|}{\textbf{\begin{tabular}[c]{@{}c@{}}Host\\ Memory\\ (Bytes)\end{tabular}}} & \multirow{2}{*}{\textbf{Status}} & \multirow{2}{*}{\textbf{Notes}} \\ \cline{2-5} 
 & \textbf{base} & \textbf{ours} & \textbf{base} & \textbf{ours} &  &  \\ \hline \hline
bt & 1459137.85 & 1582481.62 & 9059028 & 13109876 &  \cmark & Silent errors \\
cg & 19348.39 & 23165.54 & 1980816 & 3188016 & \xmark & Seg. Fault \\
ep & 162094.28 & 173420.81 & 2585996 & 3807536 & \cmark & Silent errors \\
ft & 70689.88 & 77007.16 & 2489744 & 3709232 & \cmark & Silent errors \\
is & 7786.01 & 9343.82 & 1952144 & 3166508 & \xmark & Seg. Fault \\
lu & 703200.95 & 703200.95 & 2794892 & 4004140 & \cmark & Silent errors \\
mg & 26627.67 & 30435.92 & 1975696 & 3624236 & \cmark & Silent errors \\
sp & 641277.22 & 755053.48 & 2785680 & 4000048 & \cmark & Silent errors \\ \hline
bc & 5277.73 & 5064.94 & 1878412 & 3091760 & \cmark & Silent errors \\
bfs & 254.57 & 413.62 & 1521036 & 2733360 & \cmark & Silent errors \\
cc & 399.36 & 566.8 & 1503632 & 2711856 & \cmark & Silent errors \\
pr & 6303.33 & 8070.86 & 1874316 & 3092784 & \cmark & Silent errors \\
tc & 156894.62 & 204465.25 & 2553228 & 3774764 & \xmark & Seg. Fault \\ \hline
\end{tabular}%
}
\caption{This table shows the simulation statistics of baseline gem5 and \papertitle. We also show the data corruption status (verification of the program status) of benign applications.}
\label{tab:benign-app-live}
\end{table} 

    \subsection{Additional Results}
\begin{figure}[!b]
            \centering
            \mbox {
                \subfigure[Run 1] {
                    \label{fig:real-sim-1}
                    \includegraphics[width=0.08\textwidth]{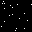}
                }
                \subfigure[Run 2] {
                    \label{fig:real-sim-2}
                    \hspace{-2ex}
                    \includegraphics[width=0.08\textwidth]{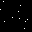}
                }
                \subfigure[Run 3] {
                    \label{fig:real-sim-3}
                    \hspace{-2ex}
                    \includegraphics[width=0.08\textwidth]{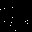}
                }
                \subfigure[Run 4] {
                    \label{fig:real-sim-4}
                    \hspace{-2ex}
                    \includegraphics[width=0.08\textwidth]{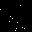}
                }
                \subfigure[Superimposed Image] {
                    \label{fig:real-sim-all}
                    \hspace{-2ex}
                    \includegraphics[width=0.08\textwidth]{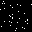}
                    }
                }
                \caption{Simulated RowHammer result of bitflips on a single row.
                This set is generated using a variation map taken from an actual
                DRAM DIMM's row (1278) (Figure~\ref{fig:rh-real-map})}
                \label{fig:rh-simulated-map-real}
            \end{figure}

        \papertitle~allows us to use statistically generated device maps.
        This generalizes the usability of the tool since we can use generic device maps based on some process variation-based probability distribution.
        Visualized results of the simulation on a specific row with 8192 columns can be seen in Figures~\ref{fig:real-sim-1} - \ref{fig:real-sim-all}.
        The hardware counterpart of the actual weak cell map is shown in Figure~\ref{fig:rh-real-map}. 
        The measured similarity index between the real and the simulated run is 0.31 in terms of JS Divergence~\cite{js-divergence} value.

\section{Related Works}
    \label{sec:related}
    RowHammer is widely studied using trace-based simulators~\cite{prohit,twice,blockhammer,graphene, panopticon}.
    The major limitation of this approach is that it does not consider 
    any of the OS effects or influences while studying both RowHammer attacks
    and defenses.
    DRAM DIMMs strictly operate based on timing.
    This led to the hypothesis that RowHammer mitigations or \textit{inhibitor}
    sends ACT commands to possible victim rows during tREFI time, when the
    device is locked for normal reads and
    writes~\cite{trrespass, blacksmith, utrr}.
    This further implies that adding victim ACTs whenever a RowHammer threshold
    is crossed may not be practical to implement by DRAM vendors.

    gem5 has been used to study architectural and micro-architectural
    vulnerabilities in the past~\cite{gem5_security,spectre_gem5}. 
    There have been attempts on implementing RowHammer at a system-level in the
    past using \gem~\cite{France2021ImplementationOR, France2021VulnerabilityAO,hammulator}.
    Most of these works use an external trace-based simulator like
    ramulator~\cite{ramulator} or DRAMsim3~\cite{dramsim3} as the backend
    memory.
    France~\textit{et al.}~\cite{France2021ImplementationOR} use
    ramulator~\cite{ramulator} to simulate the backend memory and bitflips.
    The authors also did a vulnerability assessment of the RowHammer
    attack~\cite{France2021VulnerabilityAO} using a similar setup.
    Hammulator~\cite{hammulator} uses DRAMsim3~\cite{dramsim3} as the backend
    memory.

    While the aforementioned approaches are similar to \papertitle, our technique adds the ability to experiment with different probability distributions.
    This allows further exploration of implications of RowHammer in terms of applications and experiment system-level RowHammer mitigations 
    This allows researchers to perform system-level exploits and mitigations while taking different probability distributions into account.
    
    Data corruption at a system-level allows researchers to deterministically perform RowHammer attacks and exploit system-sensitive data.
    This was demonstrated in rowhammer-test~\cite{seaborn2015exploiting} where the authors gained access to the page table from userspace by correctly flipping the access bits.
    \papertitle~improves upon these aforementioned works by adding uniqueness to the simulated DIMMs to represent real-world DIMMs via a device map.
    This not only limits \papertitle~to be a RowHammer simulator, but also
    do DRAM/memory variation studies using simulators, making them closer
    to its hardware counterpart~\cite{vampire} at a system-level.


\section{Conclusion}
    \label{sec:conclusion}

    In this work, we introduced \papertitle, a system-level framework for modeling RowHammer within the gem5 simulator.
    \papertitle~enables researchers to explore diverse RowHammer access patterns under realistic full-system conditions, including OS and cache interactions.
    By incorporating device maps, the tool captures per-DIMM vulnerability characteristics, allowing fine-grained analysis of bit-flip patterns and their implications.
    Our framework supports both synthetic attack traces and real RowHammer programs, enabling comprehensive evaluation of mitigation strategies.
    While the mitigations demonstrated in this paper are intentionally simple, \papertitle~provides an extensible foundation for prototyping and benchmarking advanced defenses.
    Looking ahead, we plan to enhance \papertitle~with richer mitigation models and introduce a quantitative RowHammer vulnerability metric at the system level, fostering rigorous and reproducible research in DRAM reliability and security.



\bibliographystyle{IEEEtranS}
\bibliography{ref_files/ref.bib}

\appendix

\subsection{Navigating the Source Code}
\label{app:using_hammersim}
    One of the objectives of this tool is to be convinient and handy to
    get started with. Towards this, we document and provide a detailed
    step-by-step instruction list.

    The repository is hosted at \href{https://github.com/darchr/gem5-rowhammer}{github}\footnote{
    \href{https://github.com/darchr/gem5-rowhammer}{https://github.com/darchr/gem5-rowhammer}.}
    The respository is hosted as fork of \gem.
    \papertitle~is encompassed within the following source files of the
    parent \gem~repository:
    \begin{itemize}
        \item \texttt{src/mem/DRAMInterface.py} All the high-level
        parameters to simulate RowHammer including RowHammer threshold,
        TRR mechanism, RowHammer probabilities \textit{etc.} are defined
        as a part of the DRAM interface. Since read/write disturbance
        follows a different distribution in the case of NVM
        devices~\cite{sttrowhammer}, we separate these parameters out and
        keep it limited to the DRAM interface in \gem.
        \item \texttt{src/mem/mem\_interface.hh} Post gem5 v.21, the memory
        interface has been split into two different interfaces to distinguish
        between DRAM and NVM devices. This file includes the necessary variables
        to keep a track of aggressor and victim rows.
        \item \texttt{src/mem/dram\_interface.hh/cc} Majority of the changes
        necessary to implement \papertitle~is incorporated here. The latter
        half of this subsection discusses more on this file.
        \item \texttt{src/mem/SConscript}
    \end{itemize}

    Another way of looking at \papertitle~is to simulate \gem~with a unique DRAM DIMM in each experiment.
    This is enabled via the \texttt{device\_file} that the user supplies in as an input parameter.
    The \texttt{device\_file} is essentially a weak cell distribution map, which we previously referred to as the device map.
    This map is generated directly from a DRAM DIMM using SoftMC~\cite{hassan2017softmc} if using an FPGA, TRRespass~\cite{trrespass}, Blacksmith~\cite{blacksmith} \textit{etc}., if using a desktop-class machine.
    Alternately this map can also be statistically generated using process variation modeling techniques like VARIUS~\cite{varius}.

    For simplicity, the map is encoded as a \texttt{.json} file with the format listed in Code Listing~\ref{code:device_map}.
    This resolution of weak cells can be tuned as per users' needs.
    A coarse-grained variation map with weak rows can model uniform distribution of a bitflip across the row, while a column or a capacitor-level map can simulate bitflips with higer fidelity with the overhead of larger variation maps.


\begin{lstlisting}[language=python, caption=Device Map Format,label=code:device_map]
{
    "rank_number": {
        "bank_number": {
            "row_number": [.., list_of_all_weak_columns, .. ],
        }
    }
}
\end{lstlisting}

    TRRs, the only practical solution to RowHammer has to ensure that
    the timing contrains of the DRAM device is consistent. Therefore, a new
    RowHammer detection mechanism has to be incorporated while the DRAM
    \texttt{ACTIVATE}s and its corresponding mitigation has to be within a
    \texttt{REFRESH}. We have implemented mitigations as \texttt{trrVariant}
    in the codebase.
    Adding a new mitigation mechanism has to be done in the \texttt{src/mem/dram\_interface.cc} file.
    This has to be within the \texttt{trrVariant} case. The Code
    Listing~\ref{code:new_mitigation} shows how to add a new
    RowHammer detection and mitigation mechanism in~\papertitle. Code
    Listing~\ref{code:check} shows where \papertitle~checks for bitflips in
    the codebase.


\begin{lstlisting}[language=C++, caption=Adding a new RowHammer Mitigation in \papertitle,label=code:new_mitigation]
// the sampler/counter mechanism is defined here.
void
DRAMInterface::activateBank(Rank& rank_ref, Bank& bank_ref, Tick act_tick, uint32_t row) {
    ...
    switch (trrVariant) {
        ...
        case N: {
            // write a new mitigation mechanism
            // here.
        }
        ...
    }
    ...
}

// the inhibitor mechanism is implemented here.
// this is because the inhibitor mechanism is 
// triggered when the DRAM device is locked for
// refreshing.
void
DRAMInterface::Rank::processRefreshEvent() {
    ...
    switch(dram.trrVariant) {
        ...
        case N: {
            // write the inhibitor mechanism
            // here to keep DRAM timing
            // consistent.
        }
        ...
    }
    ...
}
\end{lstlisting}

    From previous work, we know that the method of storing data in DRAM DIMMs is proprietary to the manufacturers~\cite{blacksmith,centauri}.
    A charged capacitor can either store a \texttt{0} or a \texttt{1}.
    Base version of gem5 does not implement this due to obvious overhead contrains.
    Therefore this is a limitation of \papertitle, as mentioned in Section~\ref{sec:limitations}.

    To implement data corruption, we selectively corrupt a random capacitor of a column.
    We assume valid data is represented as a charged capacitor in the simulation.
    \texttt{rowhammer-test}, 
    \papertitle~allows selective implementation of the same by only altering victim rows' data via the method \texttt{void doMemmoryCorruption()} thus solving the overhead issue.

\begin{lstlisting}[language=C++, caption=Checking for RowHammer Bitflips,label=code:check]
void
DRAMInterface::checkRowHammer(Bank& bank_ref, MemPacket* mem_pkt) {
...
}
\end{lstlisting}


    

        
\subsection{RowHammer Simulation Parameters}
\label{app:params}
    Table~\ref{tab:sim-params} lists all the available parameters to the user for simulating RowHammer.
    We broadly divide the parameters into five distinct categories:
    \begin{enumerate}
        \item \textit{RowHammer simulation parameters}: Parameters used to vary the probability of observing a bitflip based on a certain attack pattern.
        In addition, the user can input a RowHammer variation map as a \texttt{JSON} file as the \texttt{device\_map}.
        \item \textit{RowHammer mitigation parameters}: We have implemented two versions of TRR from prior literature~\cite{trrespass,utrr,blacksmith,blacksmith-code}.
        We preserve the \textit{sampler-inhibitor} structure~\cite{trrespass} within the DRAM interface.
        We provide several parameters to study the RowHammer sampler.
        \item \textit{Features:} We treat memory corruption as a feature in the current version of \papertitle.
        \item \textit{ECC parameters:} We add a couple of ECC parameters that the user can vary.
        This includes the ECC algorithm and its necessary input parameters for error correction (\textit{e.g.} \texttt{pMatrix} \textit{etc.}).
        \item \textit{Utilities:} We further add a couple of useful features to the user, such as dumping the TRR and bitflip information, mainly for offline \papertitle.
    \end{enumerate}

\begin{table}[t]
\centering
\resizebox{0.49\textwidth}{!}{%
\begin{tabular}{ccc}
\hline
\textbf{Category} & \textbf{Parameter} & \textbf{Description} \\ \hline \hline 
\multirow{5}{*}{\begin{tabular}[c]{@{}c@{}}RowHammer\\ simulation\\ parameters\end{tabular}} & device\_file & Path to a variation map. \\
 & rowhammer\_threshold & Standard RowHammer bitflip threshold. \\
 & single\_sided\_prob & \begin{tabular}[c]{@{}c@{}}Probability of observing a single-sided\\ RowHammer bitflip.\end{tabular} \\
 & double\_sided\_prob & \begin{tabular}[c]{@{}c@{}}Probability of observing a double-sided\\ RowHammer bitflip.\end{tabular} \\
 & half\_double\_prob & \begin{tabular}[c]{@{}c@{}}Probability of observing a half-double\\ RowHammer bitflip.\end{tabular} \\ \hline
\multirow{5}{*}{\begin{tabular}[c]{@{}c@{}}RowHammer\\ mitigation\\ parameters\end{tabular}} & trr\_variant & \begin{tabular}[c]{@{}c@{}}Mitigation variant to use during\\ simulation.\end{tabular} \\
 & trr\_threshold & \begin{tabular}[c]{@{}c@{}}Every mitigation needs to have a\\ threshold to send refreshes.\end{tabular} \\
 & companion\_threshold & \begin{tabular}[c]{@{}c@{}}Some mitigations use multiple\\ tables. This structure addresses that.\end{tabular} \\
 & counter\_table\_length & The length of the mitigation table. \\
 & companion\_table\_length & \begin{tabular}[c]{@{}c@{}}The length of the secondary counter\\ table.\end{tabular} \\ \hline
Features & enable\_memory\_corruption & Simulate live memory corruption. \\ \hline
\multirow{3}{*}{\begin{tabular}[c]{@{}c@{}}ECC\\ parameters\end{tabular}} & enable\_ecc & Binary input to enable ECC. \\
 & p\_matrix & A plain text file of a p\_matrix. \\
 & ecc\_algorithm & \begin{tabular}[c]{@{}c@{}}To switch between multiple ECC\\ algorighms.\end{tabular} \\ \hline
\multirow{3}{*}{Utilities} & trr\_stats\_dump & Dumps all TRR variable counts. \\
 & rh\_stat\_file & Dumps all bitflip variable counts. \\
 & synthetic\_traffic & To enable synthetic traffic. \\ \hline
\end{tabular}%
}
\caption{All parameters available to the user to simulate RowHammer in \papertitle.}
\label{tab:sim-params}
\end{table}


\end{document}